\documentclass[aps,physrev,twocolumn]{revtex4-2}

\usepackage{graphicx}
\usepackage{xcolor}
\definecolor{mycol}{RGB}{0,102,204} 
\usepackage[colorlinks, linkcolor=mycol, citecolor=mycol, urlcolor=mycol, breaklinks]{hyperref}
\usepackage{bbm}
\usepackage{MnSymbol}
\usepackage{physics}
\usepackage{babel}

\newcommand{\mi}{ {\rm i} }
\newcommand{\me}{ {\rm e} }

\begin{document}

\title{Disorder-induced non-Gaussian states in large ensembles of cavity-coupled molecules}

\author{R.~Schwengelbeck}
\affiliation{CESQ/ISIS (UMR 7006), CNRS and Universit\'{e} de Strasbourg, 67000 Strasbourg, France}

\author{M.~Pandini}
\affiliation{CESQ/ISIS (UMR 7006), CNRS and Universit\'{e} de Strasbourg, 67000 Strasbourg, France}

\author{R.~Daraban}
\affiliation{CESQ/ISIS (UMR 7006), CNRS and Universit\'{e} de Strasbourg, 67000 Strasbourg, France}

\author{J.~Schachenmayer}
\affiliation{CESQ/ISIS (UMR 7006), CNRS and Universit\'{e} de Strasbourg, 67000 Strasbourg, France}

\date{\today}

\begin{abstract}
We analyze vibrational dynamics in a toy model for polaritonic chemistry under collective electronic strong coupling. In a Holstein-Tavis-Cummings model, incoherently excited by a photon, we show that disorder leads to non-Gaussian states of vibrational modes on short time scales at the single-molecule level. Using exact matrix product state simulations, we demonstrate that this effect can remain robust for larger molecule numbers, implying that nuclear wave packets cannot be effectively described by thermal states. Furthermore, we compare simulations of the exact quantum dynamics with semiclassical approximations. We find that the Ehrenfest approximation can only well reproduce ensemble-averaged observables for very large system sizes. Also simulations in the truncated Wigner approximation fail to capture the non-Gaussian effects. Our work highlights the importance of disorder and genuine quantum effects in cavity-modified nuclear dynamics in polaritonic chemistry.
\end{abstract}

\maketitle

\section{Introduction}

\begin{figure*}
\includegraphics[width=0.9\textwidth]{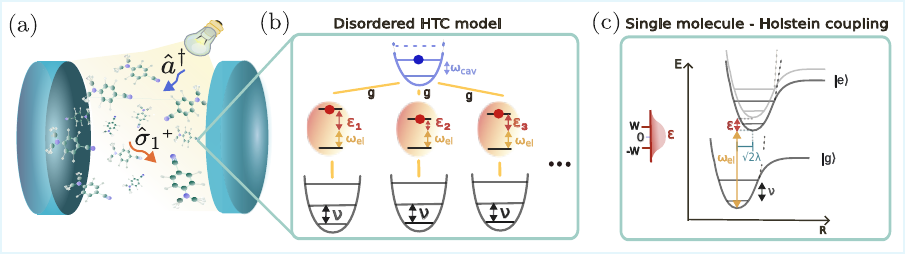}
    \caption{\textit{Schematics of our model --} (a) Setup of $N$ non-interacting molecules jointly coupled to a single optical cavity mode. The system is excited by incoherent pumping of either the photon mode or a molecular electronic transition. (b) In a toy-model setup we describe the system with a disordered Holstein-Tavis-Cummings (HTC) Hamiltonian [Eq.~\eqref{eq:htc}]: Each molecule is modeled as a single disordered two-level system (random normal distributed energy offsets $\varepsilon_i$) coupled to a single vibrational oscillator mode (frequency $\nu$). Electronic transitions couple to the cavity mode (frequency $\omega_{\rm cav}$, coupling strength $g$). (c) Electro-vibrational coupling stems from displaced vibrational potential energy surfaces for a single nuclear coordinate ($R$), approximated as harmonic oscillators near their minimum (Holstein coupling). For each molecule the vibrational potential in the electronic state $\ket{e}_i$ is shifted by $\sqrt{2} \lambda$ with respect to the one in the electronic ground state  $\ket{e}_i$.
    }
    \label{fig:sketch}
\end{figure*}

Strong coupling between molecular electronic or vibrational transitions and confined electromagnetic fields enables the formation of hybrid light-matter states, so-called polaritons. In molecular systems~\cite{feist_overview_pol_chem_2018,hertzog_strong_2019,yuen-zhou_polariton_2025,ebbciutividal}, such hybridization has been suggested to be able to reshape molecular potential energy surfaces and modify chemical reactivity. A range of experiments have reported modified reactivity under various vibrational or electronic strong coupling scenarios~\cite{ebb_el_2012,vibebb1,vibebb2, ahn2023modification}, however, microscopic details of such modifications remain only poorly understood. This has motivated significant theoretical efforts and led to the development of numerical approaches~\cite{feist_theoret_chall_pol_chem,herrera_mol_pol_for_contoll_chem_2020, Mandal} ranging from first-principle (ab initio) methods~\cite{ruggenthaler2019, bauman_perspective_2025}, to cavity quantum electrodynamics (cavity QED) model approaches~\cite{galego2015cavity, kowalewski2016non, ribeiro_polariton_2018}, as well as to semiclassical techniques~\cite{GDTWAVendrell, kanakati_benchmarking_2026}. Still, a quantitative microscopic understanding of experimental findings remains elusive.

\smallskip

A central conceptual challenge is that the coupling of molecular ensembles to cavities is collective, giving rise to polaritonic excitations being delocalized over a usually large number of $N$ molecules. In contrast, chemical processes, such as bond breaking, are governed by local nuclear coordinates and the electronic structure (e.g., the local charge distribution)~\cite{Sidler2021}.  Understanding how this collective behavior is reflected in modified local nuclear dynamics is therefore crucial to polaritonic chemistry.

\smallskip

Disorder is a promising ingredient for explaining local modification of dynamics under collective strong coupling in large ensembles~\cite{botzung2020dark,dubail2022large, sommer2021molecular, Wellnitz2022,perez2024collective,li_molecular_2026}. In this work we will in particular focus on static energetic disorder, which in experimental platforms can arise from variations in local environments. Such ``diagonal disorder'' breaks the exchange symmetry between different molecules and leads to dark states being effectively more strongly coupled to the cavity mode. In a simple Tavis-Cummings model, featuring only two-level emitters resonantly coupled to the cavity mode, diagonal disorder leads to exotic localization and transport properties of the dark states, an effect known as semilocalization~\cite{botzung2020dark, dubail2022large}. How such semilocalized states can impact the nuclear dynamics and chemical processes still remains a largely open question.

A minimal toy model for capturing nuclear dynamics in molecules under collective electronic strong coupling is the Holstein-Tavis Cummings (HTC) model~\cite{cwik2014polariton, herrera_cavitycontrolled_2016, zeb_exact_2018}. Here, nuclear dynamics are only included in the form of harmonic nuclear vibrations. Electro-vibrational coupling stems from a displaced equilibrium position of the oscillators, depending on the electronic state (Holstein coupling). The Holstein coupling leads to vibrationally-dressed molecular polaritons and to dark states acquiring a finite photon weight~\cite{herrera_cavitycontrolled_2016}. Without disorder, vibrational dynamics remains symmetric across all molecules. Diagonal disorder breaks this permutation symmetry and enhances the photon weight of the dark states. This can lead to inhomogeneous vibrational dynamics and modifications of the nuclear wave packets at the single-molecule level~\cite{Wellnitz2022}. 

For more realistic molecular descriptions, ab initio methods can describe microscopic dynamics and spectra for a few cavity-coupled molecules very precisely. However, such approaches have to make strong approximations to model collectively coupled ensembles with large $N$. In particular, semiclassical approximations can be used to treat very large system sizes and have been shown to reproduce some quantum dynamical features of the polaritonic system under collective strong coupling~\cite{GDTWAVendrell, kanakati_benchmarking_2026}. Still, it remains an open question whether such approaches can capture all relevant quantum effects in cavity-modified nuclear dynamics.

In contrast, simplified cavity QED models relying on highly simplified molecular descriptions, such as the disordered HTC model, provide an effective framework that can capture all large-scale many-body effects. Numerically exact approaches such as matrix product states (MPS) provide an exact quantum description of nonequilibrium dynamics for large systems in such models~\cite{delpino_tensor_2018,Wellnitz2022,matousek2024polaritonic,li_molecular_2026}. Previous MPS simulations of the disordered HTC model after an incoherent photo-excitation have shown that energetic disorder in the molecular transition frequencies can significantly enhance electro-vibrational entanglement and generate non-symmetric features in the vibrational wave packets~\cite{Wellnitz2022}. For individual molecules, these features were shown to decrease approximately as $1/N$, while the accumulated measure obtained by summing over all molecules remains finite in the large $N$-limit. However, a characterization of this asymmetry remains an open question, as does whether these disorder-induced signatures persist at the level of single-molecule vibrational states in the large-$N$ limit.

\smallskip

In this work, we investigate the ultrafast reduced vibrational dynamics in the disordered HTC model using numerically exact MPS simulations. We consider a situation where the system is excited by an incoherent photon, resembling scenarios relevant in polaritonic photo-chemistry. We find and quantify deviations from Gaussianity in reduced vibrational states using a non-Gaussianity measure. We show that for a single, initially excited, molecule, the local non-Gaussianity remains nearly constant as a function of $N$ in the range of $N \sim 100$, whereas in the absence of disorder it decreases with increasing molecule number. This shows that static energetic disorder can enhance the robustness of local microscopic quantum features in the HTC model, while they are suppressed at the level of ensemble-averaged vibrational states. In line with this, we find that the reduced vibrational states at short time cannot be effectively  described as thermal states. Moreover, we show that two semiclassical approximations, the truncated Wigner approximation and the Ehrenfest method, fail to reproduce the non-Gaussian features. Although the precision of such semiclassical approximations improves with $N$, deviations remain significant for systems of $N \sim 100$ molecules. 

This work is organized as follows. First, in Sec.~\ref{sec:model_section} we introduce the disordered HTC model and the dynamical scenario that we study. In Sec.~\ref{sec:non-gaussian} we introduce our measure for non-Gaussianity. There, we begin by analyzing the non-Gaussianity of the time-evolved vibrational states and investigate how it scales with the number of molecules and the disorder strength. Then, we analyze the non-thermal nature of vibrational states. Lastly, in Section~\ref{sec:semiclassical} we test the validity of the Ehrenfest and truncated Wigner approximation against the exact MPS results, and analyze the dependence of the validity on system size and disorder strength. Finally, in Sec.~\ref{sec:concl}, we provide a conclusion and an outlook.

\section{Model} \label{sec:model_section}

\subsection{Review of the disordered HTC model}
\label{ssec:htcmodel}

We study the femtosecond-scale dynamics of an ensemble of $N$ molecules coupled to a cavity [Fig.~\ref{fig:sketch}(a)], modeled by the Holstein-Tavis-Cummings (HTC) Hamiltonian~\cite{cwik2014polariton, herrera_cavitycontrolled_2016, zeb_exact_2018}. In this model, each molecule $i$ is a two-level emitter with an electronic ground state $\ket{g}_i$ and an excited state $\ket{e}_i$, separated by a bare electronic energy $\omega_{\rm el}$. Within the Born-Oppenheimer approximation, each molecule is coupled to a single quantized vibrational mode describing nuclear motion along an effective internuclear coordinate [Fig.~\ref{fig:sketch}(c)]. Additionally, the molecules are collectively coupled to a single optical cavity mode (with frequency $\omega_{\rm cav}$) in the dipole approximation [Fig.~\ref{fig:sketch}(b)]. In the resonant scenario, the Hamiltonian is given by
\begin{align}
     \hat H_{\rm HTC} = &-\lambda  \nu \sum_{i=1}^N \left(\hat b^\dagger_i + \hat b_i \right) \hat \sigma^+_i \hat  \sigma^-_i + \nu \sum_{i=1}^N \hat b^\dagger_i \hat b_i  \nonumber\\ 
     &+ \frac{g_c}{\sqrt{N}} \left(  \hat a^\dagger \sum_{i=1}^N \hat \sigma^-_i  + \hat a  \sum_{i=1}^N \hat \sigma^+_i \right) \nonumber\\
     &+ \sum_{i=1}^N \varepsilon_i \,  \hat \sigma^+_i \hat \sigma^-_i,
     \label{eq:htc}
\end{align}
where, $\hat \sigma^\pm_i = {\dyad{e}{g}_i}$ (${\dyad{g}{e}_i}$) are electronic raising (lowering) operators for molecule $i$, $\hat b_i$ are vibrational ladder operators, and $\hat a$ is the cavity annihilation operator. 

The first two terms describe the Holstein electro-vibrational coupling.  An electronic excitation of molecule $i$ displaces its nuclear equilibrium position by $\sqrt{2}\lambda$ in dimensionless oscillator units, lowering the vibrational ground-state energy by the reorganization energy $R=\lambda^2\nu$, where $\nu$ is the vibrational frequency and $\lambda^2$ is the Huang-Rhys factor. In models for cavity-coupled molecular systems, this Holstein coupling is essential to describe polaron physics~\cite{herrera_cavitycontrolled_2016,reitz2022cooperative}. The second line in Eq.~\eqref{eq:htc} describes the collective coupling of the molecules to the cavity via the Tavis-Cummings Hamiltonian, with collective coupling strength  $g_c=g \sqrt{N}$. The static energetic disorder in the last term represents unknown inhomogeneities in the molecular ensemble. The site-dependent electronic energy offsets $\epsilon_i \sim \mathcal{N}(0,W^2)$ are drawn from a normal distribution with variance $W^2$, where we denote $W$ as the disorder strength. The Hamiltonian is written in a frame rotating at the cavity frequency $\omega_{\rm cav}$, with the rotating-wave approximation applied. We focus on the resonant case, where the detuning at the Condon point $\Delta = \omega_{\rm el} + R - \omega_{\rm cav}$ vanishes.

\subsection{Dynamical framework and parameters}
\label{ssec:params}

In addition to coherent Hamiltonian dynamics, relevant incoherent processes include: incoherent pumping of the cavity or individual molecules 
(rate $\eta$), cavity photon loss (rate $\kappa$), electronic decay (rate $\gamma$), vibrational relaxation (rate $\gamma_{\rm vib}$) and thermally induced vibrational excitations. Such processes may be modeled by the Lindblad master equation, $\dot{\hat \rho} = \mathcal{L}(\hat \rho)$, with the coherent part governed by the HTC Hamiltonian and an incoherent part described by noise channels. Each incoherent channel $m$ is represented by a jump operator $\hat L_m$ and enters the dynamics through the Lindblad dissipator
\begin{equation}
     \mathcal{D}\left[ \hat L_m \right](\hat \rho) =   \hat L_m \hat \rho \hat L_m^\dagger - \frac{1}{2}\{\hat L_m^\dagger \hat L_m, \hat \rho \},
\end{equation}
multiplied by its corresponding incoherent rate, e.g. $\eta, \kappa, \gamma, \gamma_{vib}$. In this work, we focus on ultrafast dynamics within a single vibrational oscillation period, $0 \leq t \leq 2\pi/\nu$. On this time scale, we neglect incoherent rates for the parameter regime discussed below. We use parameters representative of typical photo-chemistry setups with organic dye molecules, following our previous work~\cite{Wellnitz2022}. For the electronic transition, we consider parameters corresponding to the organic dye molecule Rhodamine 800, with $\omega_{el} \sim 2\,$eV, for which Rabi splittings of the order of eV have been observed~\cite{Strongcouplingbetweensurfaceplasmonpolariton}. We choose a Rabi splitting of $\Omega_{\rm el} = 2g_c = 700\,$meV, consistent with reported polaritonic chemistry setups with electronic strong coupling~\cite{ebb_el_2012}.

The vibrational frequency is set to $\nu = 0.3 g_c = 105\,$meV, and the electron-vibration coupling is characterized by a Huang–Rhys factor of $\lambda = 0.4$, obeying $\lambda \nu \le \nu \le g_c$ and within the range observed for Franck-Condon active vibrational modes in Rhodamine 800~\cite{christensson2010}. These values place the system in the strong coupling regime, $\gamma ,\kappa \ll  g_c \ll \omega_{\rm el}, \omega_{\rm cav}$, with $g_c/\omega_{\rm el} = 0.18 $, so that the rotating wave approximation underlying the Tavis-Cummings Hamiltonian remains well justified.

All incoherent rates lie well below the vibrational frequency $\nu$.
Electronic decay is extremely slow, $\gamma = 10^{-7}-10^{-6}$   meV~\cite{herrera_theory_2018}, modeled by lowering operators $\hat L_{\gamma , i} = \hat \sigma_i$.
Vibrational relaxation of the nuclear motion occurs with rates $\gamma_{\rm vib, \downarrow} = 0.07-0.7$ meV~\cite{herrera_theory_2018}, with thermal excitation suppressed to $\gamma_{\rm vib,\uparrow} = \gamma_{\rm vib,\downarrow}\exp(-\nu/k_B T) \sim 10^{-3}\text{--}10^{-2}\,$meV at room temperature; these processes are described by $\hat{L}_{i,\downarrow} = \hat{b}_i$ and $\hat{L}_{i,\uparrow} = \hat{b}_i^\dagger$, respectively.
Cavity photon loss, described by $\hat{L}_{\rm cav,\kappa} = \hat{a}$ is reported in Fabry-Perot microcavities formed by distributed Bragg reflectors (DBR) to range from approximately $\sim 0.002 $ to $2$ meV~\cite{bhuyan_rise_2023}.

We assume weak incoherent driving at rates $\eta < 10^{-2}$ meV, modeled either as incoherent pumping of a cavity mode  ($\hat L_{\rm cav, \eta} = \hat a^\dagger$) or as an individual molecule excitation ($\hat L_{{\rm mol}, \eta, i} = \hat \sigma^\dagger_i$). 
For example, consider a household light source of $5$--$10\,$W placed at a distance of $0.1\,$m from a cavity with a mode area of order $1\,\mu\mathrm{m} \times 1\,\mu\mathrm{m}$. Only a tiny fraction of the emitted light intersects the cavity, yielding an effective incoherent excitation rate that we then estimate on the order of $\eta \sim 10^{-3}\,$meV. For weaker sources such as candlelight (${\sim}\,1\,$W), this rate would drop further to $\eta \sim 10^{-4}\,$meV.

\smallskip

All discussed incoherent processes thus occur at rates below the characteristic vibrational oscillation scale $\gamma, \gamma_{\rm vib}, \kappa, \eta \ll \nu$. This justifies a description where we only consider coherent dynamics at the ultrafast time scale, after excitation with a single photon (throughout, we set $\hbar \equiv 1$),
\begin{equation}
    \frac{d \hat \rho}{dt} = - \mi \left[ \hat H_{\rm HTC}, \hat \rho\right] .
\end{equation}
We consider two initial conditions within the single-excitation manifold: incoherent excitation of the cavity mode or of a single molecule. Note that we employ a density matrix description to account for the unknown static energetic disorder introduced in section~\ref{ssec:htcmodel}. Each disorder realization $k$ corresponds to a distinct Hamiltonian $\hat H^{[k]}_{\rm HTC}$ and thus to a different unitary evolution. Starting from the same initial state, we simulate the unitary dynamics for each disorder configuration separately and construct physical observables from the disorder-averaged density matrix. Although each individual time evolved state $\hat \rho(t)$ remains pure, the averaging over different unitary evolutions yields a mixed state
\begin{equation}
    \overline{\hat \rho(t)} = \frac{1}{N_D} \sum_k^{N_D} \hat \rho^{[k]}(t).
\end{equation}
This approach is an appropriate description of systems with static disorder when the specific disorder configuration is unknown and varies slowly compared to the time scale between incoherent excitation events. 
In between photo-excitations, we assume that the slower incoherent process thermalizes the state, noting that at room-temperature the vibrational ground state is an excellent approximation to the equilibrium state, e.g.,~with a  thermal population probability of the first excited vibrational state less than $2\%$ ($\nu = 100$ meV, $k_BT \approx 25$meV). In the following, we analyze the resulting vibrational dynamics in the states $\overline{\hat \rho(t)}$ and for individual disorder realizations on the ultrafast scale.

\section{Non-Gaussian States}
\label{sec:non-gaussian}

To analyze quantum features in the ultrafast vibrational dynamics, we begin by defining and computing non-Gaussianity and its scaling with disorder strength and system size in Sec.~\ref{ssec:nongauss}. Then, we show that resulting non-Gaussian states cannot be properly described with effective thermal distributions in Sec.~\ref{ssec:therm}.

\subsection{Non-Gaussian vibrational states}
\label{ssec:nongauss}

To evaluate the impact of disorder on vibrational dynamics, we simulate the exact quantum dynamics of the disordered HTC Hamiltonian using matrix product state (MPS) techniques as introduced in~\cite{Wellnitz2022}. As argued in Sec.~\ref{ssec:params}, for our parameter regime, we consider disorder-averaged unitary dynamics of trajectories after rare incoherent excitation events of the equilibrium state, acting either on the cavity mode (with $\hat L_{{\rm cav,\eta}}$), or a single molecule (with $\hat L_{{\rm mol},\eta, i}$).  This leads to the two different normalized initial states $|\Psi_{0}^{\rm cav}\rangle = \hat a^\dagger |0\rangle_{\rm cav} |0\rangle_{\rm el} |0\rangle_{\rm vib}$, or $|\Psi_{0}^{\rm mol} \rangle = \hat \sigma^+_1 |0\rangle_{\rm cav} |0\rangle_{\rm el} |0\rangle_{\rm vib}$, respectively. 
Here w.l.o.g.~we choose the first molecule ($i=1$) as the excited one.
We select $N_D$ different static electronic disorder realizations.
For each realization $k$, the time-dependent reduced vibrational density matrix of molecule $i$ is obtained as
\begin{equation}
    \hat \rho_i^{[k]}(t) = {\rm tr}_{\neg i}\left( \ket{\psi^{[k]}(t)}\bra{\psi^{[k]}(t)}\right),
\end{equation}
where ${\rm tr}_{\neg i}$ denotes the partial trace over all degrees of freedom except the vibrational mode of molecule $i$.
The disorder-averaged vibrational state of the molecule is then obtained by computing
\begin{equation}
   \overline{ \hat \rho_i(t)}= \frac{1}{N_D}\sum_{k=1}^{N_D} \hat \rho_i^{[k]}(t).
\end{equation}
Depending on the excitation scenario, we focus on two types of reduced states. For an initial molecular excitation, we are interested in the state of the specifically excited molecule ($i=1$)
\begin{equation}
     \hat \xi_1 = \overline{ \hat \rho_1(t) }.
\end{equation}
At the same time, for both types of excitations, we consider also the ensemble-averaged vibrational state
\begin{equation}
    \hat{\xi}_{\rm avg} = \frac{1}{N}\sum_{i=1}^N  \overline{ \hat \rho_i(t) }.
    \label{eq:xi-avg-def}
\end{equation}
The quantity $\hat \xi_1$ provides us with insight on the cavity-modified dynamics of a single incoherently excited molecule, while $\hat{\xi}_{\rm avg}$ estimates the overall impact of a single excitation on the ensemble of molecules on average.

To visualize the vibrational dynamics in phase space, we expand the reduced vibrational density matrix in the Fock basis 
\begin{equation}
    \hat \xi_{1/ \rm avg} = \sum_{nm} c^{1/{\rm avg}}_{nm} |n\rangle \langle m|,
    \label{eq:xi_fs}
\end{equation}
and compute the corresponding Wigner function via
\begin{equation}
       W_{1/\rm avg}(x,p) = \frac{1}{\pi} \sum_{nm} c^{1/{\rm avg}}_{nm} \int_{-\infty}^\infty \psi_n(x-y) \psi_m^*(x+y) e^{i2py  } {\rm d}y. \label{eq:wigner}
\end{equation}
Here, $\psi_n(x)$ denotes the harmonic oscillator eigen\-functions. 

We quantify the non-Gaussian character of a state using the quantum relative entropy \cite{quantum_rel_entropy}. For a given density matrix $\hat \rho$, this quantity measures the entropy difference between  $\hat \rho$ and its closest Gaussian state $\hat \tau$,
\begin{equation}
    \delta[\hat \rho] = S(\hat \tau) - S(\hat \rho),
    \label{eq:delta-def}
\end{equation}
where $S(\hat \rho) = -\rm{tr}\left( \hat \rho \ \rm{ln}\hat \rho \right)$ denotes the von Neumann entropy. The reference state $\hat \tau$ is defined as the unique Gaussian state whose first and second moments are identical to those of the state $\hat \rho$. As a Gaussian state, $\hat \tau$ is thus fully characterized by the covariance matrix $\hat V$ of the state $\hat \rho$ \cite{Weedbrook2012}. The entropy of $\hat \tau$ can then be computed directly from $\hat V$, since it depends solely on the symplectic eigenvalue of the covariance matrix \cite{Weedbrook2012,demarie2012pedagogicalintroductionentropyentanglement,juarez_quantum_2023}. Details on the computation of the Gaussian reference state and its von Neumann entropy are given in Appendix \ref{appendix-tau-closest-gaussian}.

\medskip

\begin{figure*}
    \centering
    \includegraphics[width=0.8\linewidth]{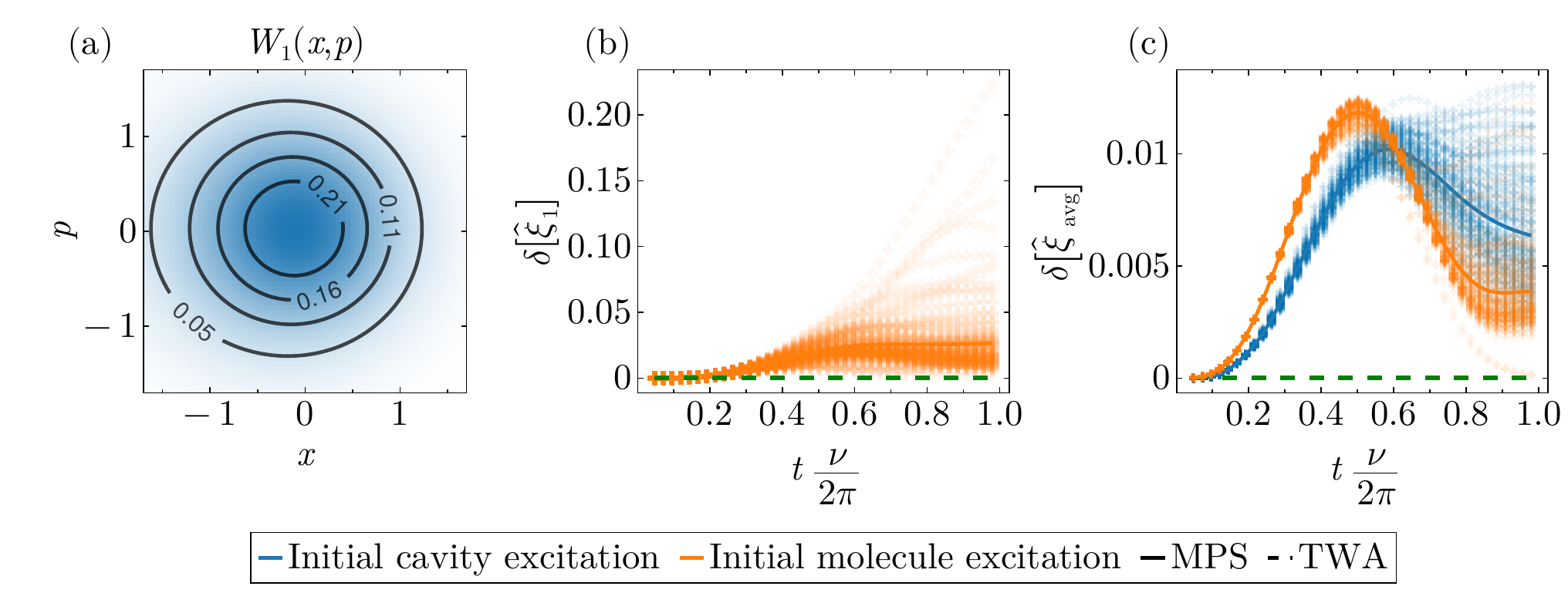}
    \caption{\textit{Non-Gaussian features --} (a) Wigner function of the disorder-averaged vibrational state $\hat \xi_1$ of the initially excited molecule after one vibrational oscillation ($t=2\pi/\nu$). (b/c) Time-evolution of the non-Gaussianity after incoherent molecule (orange) or cavity excitation (blue). Markers show individual disorder realizations, lines the mean value over all  disorder realizations. Panels (b) and (c) show results for the initially excited molecule and the ensemble average, $\delta[\hat \xi_1]$ and $\delta[\hat \xi_{\rm avg}]$, respectively. While solid lines and markers show exact results (MPS simulations), dashed lines show semiclassical TWA simulations incapable of capturing the finite non-Gaussianity. Here, we consider $N=50$ molecules, disorder strength $W=0.5g_c$, $N_D=150$ disorder realizations, $\nu = 0.3 g_c$, and $\lambda=0.4$. MPS simulations are converged with $\chi=50$ and $n_{\rm max}^{\rm vib} = 8$.}
    \label{fig:somefeatures}
\end{figure*}

Fig.~\ref{fig:somefeatures} summarizes the emergence of weak non-Gaussian features in the dynamics of a single vibrational mode for a system with $N=50$ molecules and disorder strength $W/g_c=0.5$. In Fig.~\ref{fig:somefeatures}(a), we display the Wigner function of the reduced state $\hat \xi_1$, after one vibrational period, $t = {2\pi}/{\nu}$. The non-circular contour lines show that it has a weakly asymmetric shape, which does not reflect squeezing but rather non-Gaussianity, as quantified by finite $\delta[\hat\xi_1] > 0$.  

\begin{figure*}
    \centering
    \includegraphics[width=0.9\textwidth]{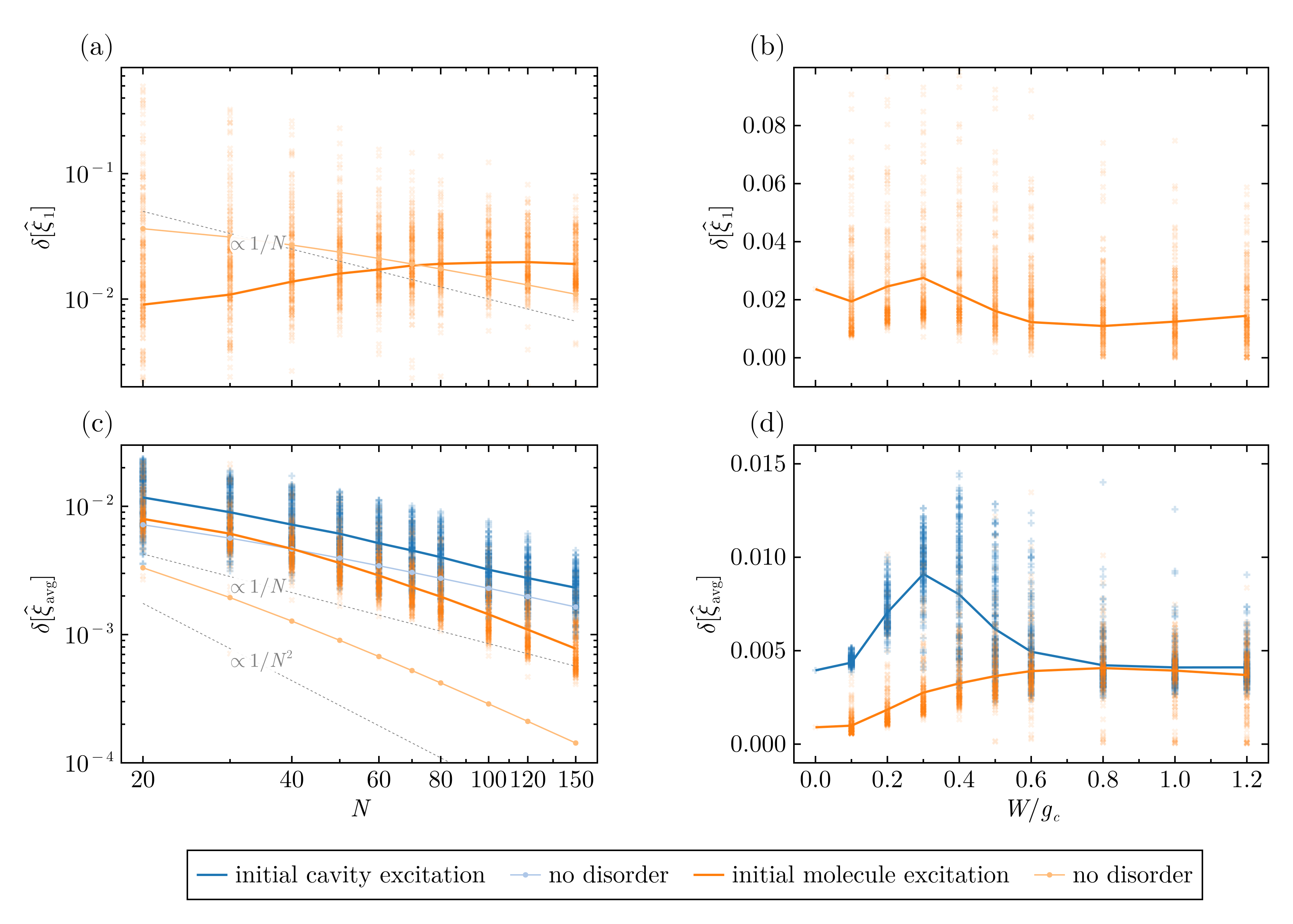}
    \caption{\textit{Non-Gaussianity scaling after one vibrational period $t = 2\pi/\nu$} -- (a,b) Non-Gaussianity of the single molecule that is initially excited ($\delta[\hat \xi_1]$). (a) $\delta[\hat \xi_1]$ as a function of molecule numbers $N$ for a fixed disorder strength of $W = 0.5 g_c$. (b) $\delta[\hat \xi_1]$ as a function of disorder strengths $W$ for a fixed number of $N=50$ molecules. (c,d) The same plots for the site-averaged non-Gaussianities $\delta[\hat\xi_{\rm avg}]$. Shown are results for both initial cavity (blue) and molecular (orange) excitation scenarios. Markers: individual disorder realizations; solid lines: disorder averages. Light lines with circular markers: disorder-free case. Dotted gray lines indicate power laws (double-logarithmic scale) as guide to the eye. Parameters:  $N_D=200$ disorder realizations, $\nu = 0.3 g_c$, $\lambda=0.4$. Exact MPS results, converged with $\chi\leq 128$ and $n_{\rm max}^{\rm vib} = 8$.
    }
    \label{fig:delta-scaling}
\end{figure*}

The non-Gaussianity of an initially excited molecule, $\delta[\hat\xi_1]$, is strongly sensitive to disorder during its time-evolution, as shown in Fig.~\ref{fig:somefeatures}(b). Individual disorder realizations $\delta[\hat\xi_1^{[k]}]$ exhibit a broad spread around the disorder-averaged curve, indicating that the local vibrational dynamics depend significantly on the specific disorder configuration. Crucially, below our key finding will be that $\delta[\hat\xi_1]$ remains finite at large system sizes $N$ (Fig.~\ref{fig:delta-scaling}), in contrast to the ensemble-averaged case.

The ensemble-averaged non-Gaussianity $\delta[\hat\xi_{\rm avg}]$ displays qualitatively different behaviors [Fig.~\ref{fig:somefeatures}(c)]. Following initial molecular excitation, it peaks sharply at half the vibrational period, $t = \pi/\nu$; for initial cavity excitation, the peak shifts slightly to $t \gtrsim \pi/\nu$. The sensitivity to individual disorder realizations is weak at short times but grows at later times. Importantly, we will see that the overall magnitude of $\delta[\hat\xi_{\rm avg}]$ decreases with increasing $N$ and vanishes in the thermodynamic limit, as analyzed in detail in Fig.~\ref{fig:delta-scaling}. Furthermore, we note that already for $N=50$, $\delta[\hat\xi_{\rm avg}]$ is an order of magnitude smaller than $\delta[\hat\xi_1^{[k]}]$.

In both cases in Fig.~\ref{fig:somefeatures}(b,c), the semiclassical truncated Wigner approximation fails to reproduce the non-Gaussian features, as shown by the flat green dashed lines. This underscores the role of quantum correlations in vibrational dynamics, consistent with Ref.~\cite{Wellnitz2022}, which attributed asymmetries of the nuclear wave packet to disorder-induced entanglement between the electro-photonic and vibrational Hilbert spaces.

\medskip

We now address the important question of whether this non-Gaussianity can persist in the large-$N$ limit. Fig.~\ref{fig:delta-scaling} shows how $\delta[\hat \xi_{1}]$ and $\delta[\hat \xi_{\rm avg}]$ scale with (a/c) system size $N$ and (b/d) disorder strength $W$.

We start with analyzing the non-Gaussianity of a single initially excited molecule, $\delta [\hat\xi_1]$ in Fig.~\ref{fig:delta-scaling}(a). In the disorderless case (light solid line), as expected (see discussion below), we observe that $\delta[\hat \xi_{1}]$ decreases with $N$, for large $N$, following approximately a $1/N$ scaling behavior. In strong contrast, the presence of disorder (solid lines and scattered markers) makes the presence of finite $\delta[\hat \xi_{1}]$ much more robust in the large-$N$ limit. We observe an initial increase of $\delta[\hat \xi_{1}]$ with $N$, with $\delta[\hat \xi_{1}]$ overcoming the value of the disorder-free case and saturating at a constant value for $N \gtrsim 80$.

We can analytically understand this behavior considering only the electro-photonic dynamics of a disordered Tavis-Cummings model, by making the assumption that the scaling of $\delta[\hat \xi_{1}]$ should follow roughly the scaling of the ``escape probability'' of an excitation from the initially excited molecule, $1 - \bra{\psi(t)}\hat \sigma_1^+ \hat \sigma_1^-\ket{\psi(t)}$. For the latter, simple perturbative expressions can be derived for the strong-coupling regime with $g_c \gg W$~\cite{botzung2020dark, dubail2022large, Wellnitz2022}. For the disorderless case $W=0$, it is straightforward to compute that the dynamics (involving only the bright collective polariton states) leads to $1- \bra{\psi(t)}\hat \sigma_1^+ \hat \sigma_1^-\ket{\psi(t)} = 2 \cos(g_ct - 1)/N + \mathcal{O}(1/N^2)$~\cite{Wellnitz2022}. This implies that the modification of the cavity on the excitation transfer away from the initial molecule vanishes as $1/N$, which explains the absence of any potential cavity-induced non-Gaussianity in the HTC model and the scaling behavior for large $N$ and $W=0$.

Remarkably, also in the presence of weak disorder, a perturbative argument leads to a scaling of $1- \bra{\psi(t)}\hat \sigma_1^+ \hat \sigma_1^-\ket{\psi(t)} \sim Wt/N$~\cite{Wellnitz2022}, suggesting a vanishing effect of the cavity, also scaling as $1/N$. This is in contradiction to our numerically exact results in Fig.~\ref{fig:delta-scaling}(a). We point out that our numerical calculations are outside the perturbative regime with $W = 0.5 g_c$. Furthermore, the perturbative results from~\cite{botzung2020dark, dubail2022large, Wellnitz2022} rely on a uniform box distribution with $\varepsilon_i \in [-W/2,W/2]$ that excludes the tails of the normal disorder distribution considered here. We consider the numerically found non-perturbative scaling of $\delta[\hat \xi_{1}] \sim \text{const.}$ in Fig.~\ref{fig:delta-scaling}(a) a key finding of our work, nevertheless, it is important to point out that our numerical calculations are limited to $N \leq 150$. Also, the scaling can strongly depend on the observable, as seen e.g.,~in comparison with the results for the asymmetric shapes of nuclear distributions in~\cite{Wellnitz2022}, and e.g.,~in comparison with the results of Fig.~\ref{fig:overlap-scaling} below. Additionally, for observing robust modifications of cavity-induced molecular dynamics in the large $N$ limit, here $W$ needs to effectively increase with the number of coupled molecules to conserve the condition $W = 0.5 g_c$, which may be an unrealistic assumption for real experiments. Nevertheless, this numerical finding highlights the importance of disorder leading to a more robust cavity-modified dynamics, beyond perturbative limits.

The ensemble average $\delta[\hat \xi_{\rm avg}]$ in Fig.~\ref{fig:delta-scaling}(c) decreases with $N$ for both initial excitation scenarios. Over the numerically accessible range of system sizes $N$, this decrease follows an approximate power law with exponents comparable to the disorder-free case, but with modified prefactors. Thus, disorder affects the overall magnitude of averaged non-Gaussianity, without changing its scaling.

The faster decay for initial molecular excitation reflects the localized nature of the excitation. Since only a single site carries non-Gaussian features, its relative contribution scales as $\propto 1/N$.  In contrast, an initial cavity excitation couples collectively to all molecules, resulting in a comparatively slower decay with system size $N$ at the short time scale considered. In the cavity excitation case, here $\delta[\hat \xi_{\rm avg}]$ follows a $\sim 1/N$ scaling. We note that since $N$ molecules all exhibit on average this non-Gaussianity, the overall ``summed-up non-Gaussianity'' remains constant also for large $N$. This is in agreement with the findings for the overall asymmetric shape observed in \cite{Wellnitz2022}. It highlights again the fact that for an initial incoherent cavity excitations, robust modifications are observable, also for large $N$, especially in the case of sufficiently large disorder.

\smallskip

Finally, we also analyze the dependence of the non-Gaussian features on disorder strength $W/g_c$ in Fig.~\ref{fig:delta-scaling}(b,d). The single-molecule non-Gaussianity $\delta [\hat\xi_1]$ (b), here for $N=50$ remains approximately nearly invariant as a function of $W$, exhibiting a maximum around $W/g_c = 0.3$. An increasing non-Gaussianity in the  strong-coupling regime $W \ll g_c$ is plausible from our perturbative argument, from which we expect the scaling $ \sim Wt/N$. The decreasing value of  $\delta [\hat\xi_1]$ for $W/g_c \gtrsim 0.3$ thus already indicates an invalidity of the perturbative strong-coupling regime. We note again that the non-Gaussianities strongly fluctuate with different disorder realizations.

The ensemble-averaged non-Gaussianity $\delta[\hat\xi_{\rm avg}]$ for the initial cavity excitation ($N = 50$) displays a pronounced maximum near $W/g_c \sim 0.3$ [Fig.~\ref{fig:delta-scaling}(d)]. This can again be understood from a perturbative argument in the disordered Tavis-Cummings model~\cite{Wellnitz2022}: asuming that cavity-mediated effects scale with the total photon weight of the dark states, the latter can be found to scale as $\propto W/g_c$ at weak disorder. The subsequent decrease of $\delta[\hat\xi_{\rm avg}]$ for $W/g_c \gtrsim 0.3$ signals again breakdown of this perturbative picture already at moderate disorder strengths. For initial molecular excitation, $\delta[\hat\xi_{\rm avg}]$ is considerably smaller and saturates at large $W$, consistent with the localized nature of the excitation. In the limit $W \gg g_c$, all cavity-induced effects have to vanish; it is therefore notable that disorder-induced non-Gaussian features remain relatively robust even at disorder strengths well outside the strong-coupling regime.

\subsection{Non-thermal vibrational states} 
\label{ssec:therm}

In this section we address the question whether the short-time evolved vibrational subsystem states may be effectively approximated as being thermal? Standard reaction rate theories often assume that internal molecular degrees of freedom equilibrate rapidly, and that the dynamics remains in thermal equilibrium throughout reaction processes~\cite{Eyring1935, Avery1974}. Then, temperature becomes the single controlling parameter.
Naturally, in our non-equilibrium model, the non-Gaussianity observed in Sec.~\ref{ssec:nongauss} already signals departure from thermal equilibrium~\cite{Weedbrook2012}.
Nevertheless, we can ask whether thermal states can effectively still approximate the time-dependent reduced density matrices $\hat{\xi}_1$ and $\hat{\xi}_{\rm avg}$, which would allow an efficient characterizations of cavity-modifications through a single effective temperature.
This expectation is particularly motivated for the strongly disordered case, where disorder-enhanced entanglement between vibrational and electro-photonic degrees of freedom~\cite{Wellnitz2022} could drive effective thermalization.

\begin{figure*}
    \includegraphics[width=1\textwidth]{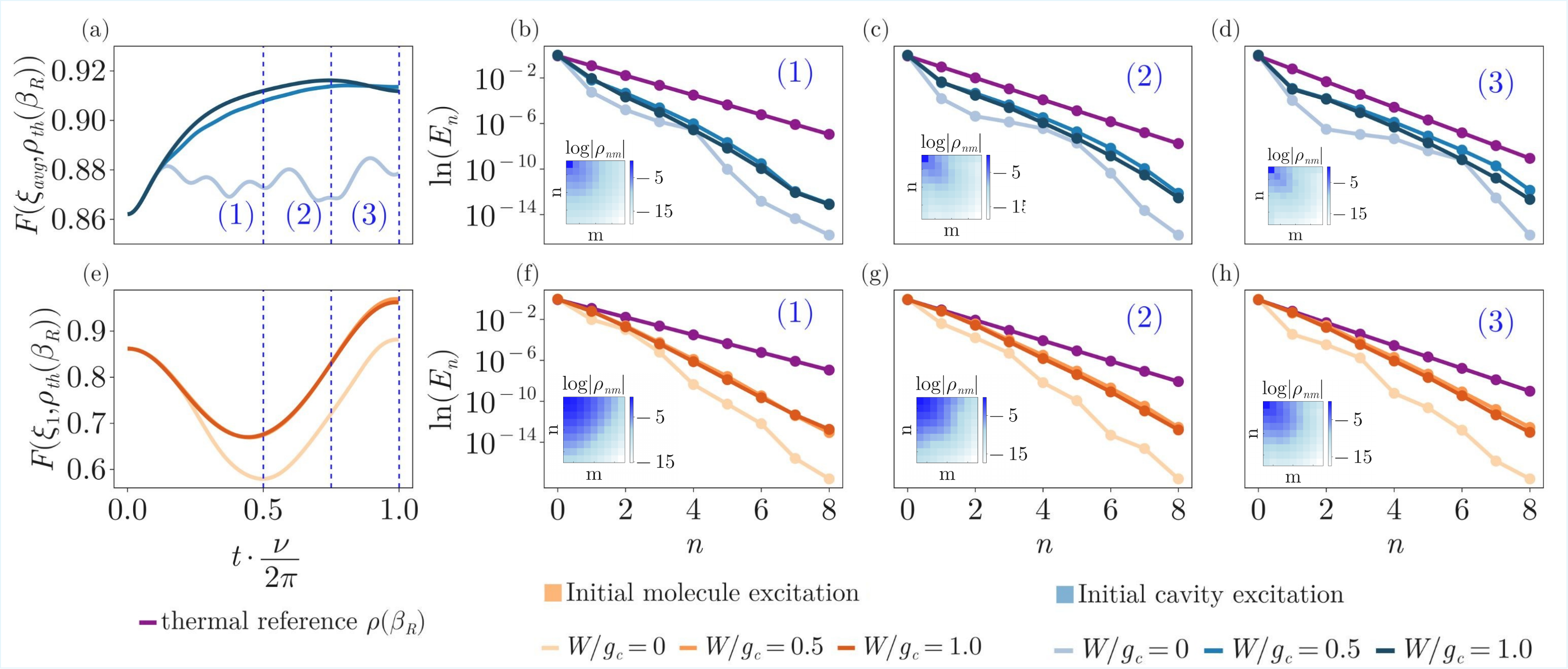}
    \caption{\textit{Comparison with thermal states --} (a–d) Initial molecular excitation, (e–h) initial cavity excitation, averaged over all molecules. (a) and (e) show the fidelity between the time-evolved vibrational state and the thermal reference for three disorder strengths $W/g_c = 0, 0.5, 1.0$. (b–d) and (f–h) display the eigenvalues of the vibrational density matrices $\hat \xi_1$ and $\hat \xi_{\rm avg}$, respectively, on a logarithmic scale at three representative times labeled (1), (2), and (3) in (a). The eigenvalue distribution indicates how close the density matrix is to a Boltzmann distribution in its eigenbasis (straight line). The purple curve shows the populations of the thermal reference state with the physically motivated inverse temperature $\beta_R$. Heatmap insets show the logarithm of absolute values of the density matrix elements, $|\langle n|\hat \xi_{1/\rm avg}|m\rangle|$ in the Fock basis for $W = 0.5 g_c$. Parameters: $N=50$, $N_D=100$ disorder realizations. Exact MPS results, converged with $\chi=64$ and $n_{\rm max}^{\rm vib} = 8$.}
    \label{fig:thermal}
\end{figure*}

\medskip

We define a thermal reference state in the Fock basis $\{ \ket{n}  \}$ of a single-molecule vibrational harmonic oscillator (setting the ground-state at zero energy) as
\begin{equation}
    \hat \rho_{\rm th} = \frac{1}{Z} \sum_n e^{-\beta \nu n} |n\rangle \langle n|,
\end{equation}
with $\beta$ the inverse temperature, and $Z= \sum_n e^{-\beta \nu n} = 1/(1-\me^{-\beta \nu})$ the partition function. We fix an inverse temperature $\beta_R$ by requiring that the thermal state reproduces the reorganization energy associated with the sudden initial excitation. This yields
\begin{equation}
    \beta_R =  \frac{1}{ \nu} {\rm ln}\left(1+ \frac{ \nu}{E_0}\right),
    \label{eq:betaR}
\end{equation}
where $E_0$ is set by the reorganization energy $R=\lambda^2 \nu$ (section \ref{ssec:htcmodel}) originating from the Holstein coupling. For an initially excited molecule, the vibrational potential is displaced, transferring an energy of $R$ to the vibrational mode; accordingly, we choose $E_0 = R$. For initial cavity excitation, we assume that this excitation is collectively distributed over all $N$ molecules. Therefore, in this case we take $E_0 =R/N$.

\smallskip

In Fig.~\ref{fig:thermal} we compare the features of the time-evolved vibrational density matrices $\hat \xi_{1/ \rm avg}$ to those of $\hat \rho_{\rm th}$. In particular, we analyze: i) the overlap of $\hat \xi_{1/ \rm avg}$ with $\hat \rho_{\rm th}$, quantified by the fidelity
\begin{equation}
    F(\hat \xi_{1/ \rm avg}, \hat \rho_{\rm th}) = \left(\rm{tr}\sqrt{\sqrt{\hat \rho_{\rm th } }\hat \xi_{1/ \rm avg} \sqrt{\hat \rho_{\rm th } }} \right)^2,
\end{equation}
ii) the eigenspectra of $\hat \xi_{1/ \rm avg}$ and $\hat \rho_{\rm th}$, and iii) the off-diagonal elements of $\hat \xi_{1/ \rm avg}$. 

\smallskip

In Fig.~\ref{fig:thermal}(a) and (e) we plot the time-evolution of $ F(\hat \xi_{1/ \rm avg}, \hat \rho_{\rm th}(\beta_R))$ for (a) the initial cavity excitation and the ensemble-averaged density matrix $\hat \xi_{\rm avg}$; and (e) for the initial molecular excitation and the single-molecule density matrix $\hat \xi_1$. We consider a moderate system size of $N=50$. 

In both excitation scenarios we consider disorder strengths of $W/g_c=\{0, 0.5, 1.0\}$. For the cavity excitation, in Fig.~\ref{fig:thermal}(a) we observe that for all disorder strengths, the fidelity quickly saturates at a value of around $90-95\%$ in the first oscillation period. This implies that the vibrational states do not approach the thermal state $\hat \rho_{\rm th}(\beta_R)$ with the evolution on our ultrafast scale $0 \leq t \leq 2\pi/\nu$. For the single-molecule excitation, in Fig.~\ref{fig:thermal}(e) we observe much lower fidelities of the excited molecule, reaching to values as low as $60\%$ at around half the oscillation period. 

These observations generally correlate with the non-Gaussianity in Fig.~\ref{fig:somefeatures} and Fig.~\ref{fig:delta-scaling}: i) $\delta\left[ \hat \xi_1 \right]$ after a molecular excitation remains significantly larger than $\delta\left[ \hat \xi_{\rm avg} \right]$ after cavity-excitation, and the same is true for the fidelity; ii) The largest infidelities and non-Gaussianities both occur at around half an excitation period. However, interestingly, in both excitation scenarios the disorderless case $(W=0)$ clearly exhibits the smallest fidelity with the thermal reference state, while the corresponding non-Gaussianity in Fig.~\ref{fig:delta-scaling}(b) has very similar magnitudes for all the considered cases of $W/g_c = \{0, 0.5, 1\}$. A possible explanation for this is given by the fact that disorder increases the entanglement entropy between the vibrational and the electro-photonic part of the wavefunction in single trajectories~\cite{Wellnitz2022}. Comparing these previous findings to our results in Fig.~\ref{fig:thermal}, we conclude that this entropy can contribute to a more thermal behavior. However, importantly, we do find that the density matrices on the short time scale still remain significantly different from $\hat \rho_{\rm th}$.

So far we have examined the overlap of the density matrix with a thermal state at fixed inverse temperature $\beta_R$. However, it could also be the case that the time-dependent states resemble thermal states at some other temperature $\beta^*$. We examine this scenario, by displaying the eigenspectra of $\hat \xi_{\rm avg}$ and $\hat \xi_{1}$, as well as of $\hat \rho_{\rm th}$ in Fig.~\ref{fig:thermal}(b-d) and (f-h), for the two different excitation scenarios, respectively. The different panels (b)/(f), (c)/(g), and (d)/(h) correspond to the different times $t = \pi/ \nu$, $t = 3\pi/ 4\nu$, and $t = 2\pi/ \nu$, as marked by the vertical dashed lines in (a/e), respectively.

A thermal state corresponds to a diagonal density matrix in the Fock basis with the diagonal values following a Boltzmann distribution (line with slope $-\beta_R$ on the natural log-scale). Instead, we note that at all times and all excitation scenarios the time-dependent reduced density matrices are not diagonal in the Fock basis. This is seen in the different insets, which show the logarithm of the absolute values of the density matrix elements $|\langle n|\hat \xi_{1/ \rm avg}|m\rangle|$ in the Fock basis for $W=0.5g_c$. They all feature significant off-diagonal contributions. 

Furthermore, we plot the eigenvalues of $\hat \xi_{1/ \rm avg}$ in each panel for different disorder strengths of $W/g_c = \{0, 0.5, 1\}$, matching colors in (a) and (e). The plots reveal that the spectra of $\hat \xi_{1/ \rm avg}$ do not approach the one of the thermal reference $\hat \rho_{\rm th}$, nor the shape of another effective thermal Boltzmann distribution: For initial cavity excitation, the disorderless case $W = 0$ deviates most strongly from an exponential decay. The same trend is observed for the initial molecular excitation. For both finite disorder strengths $W/g_c = \{0.5, 1\}$, the eigenvalue distributions are very similar to each other, in both excitation scenarios. They are closer to a thermal distribution, which is plausible given that disorder enhances the entanglement between the vibrational subsystem and the rest~\cite{Wellnitz2022}.  However, it is important to note that in both cases the eigenspectra do not feature a clear exponential decay for either of the analyzed points in time. Instead, in the case of cavity excitation (b-d), we observe a drop from the largest to the second-largest eigenvalue. For the initial molecule excitation (f-h) we observe a general ``bending-down'' of the spectrum indicating a super-exponential distribution. 

\smallskip

This leads us to the conclusion that on the short time scale after incoherent excitation, the Hamiltonian dynamics does not lead to vibrational states that can be modeled by an effective thermal density matrix. Furthermore, we point out that on our time scales of interest the eigenvalue distribution of $\hat \xi_{1/ \rm avg}$ consistently exhibit a faster decay than those of the thermal reference state $\hat \rho_{\rm th}$. Therefore, if one was to assign an effective temperature $1/\beta^*$ to the vibrational states shortly after incoherent excitation, this effective temperature would remain smaller than $1/\beta_R$ estimated from Eq.~\eqref{eq:betaR}.

\begin{figure*}
    \centering
    \includegraphics[width=0.9\textwidth]{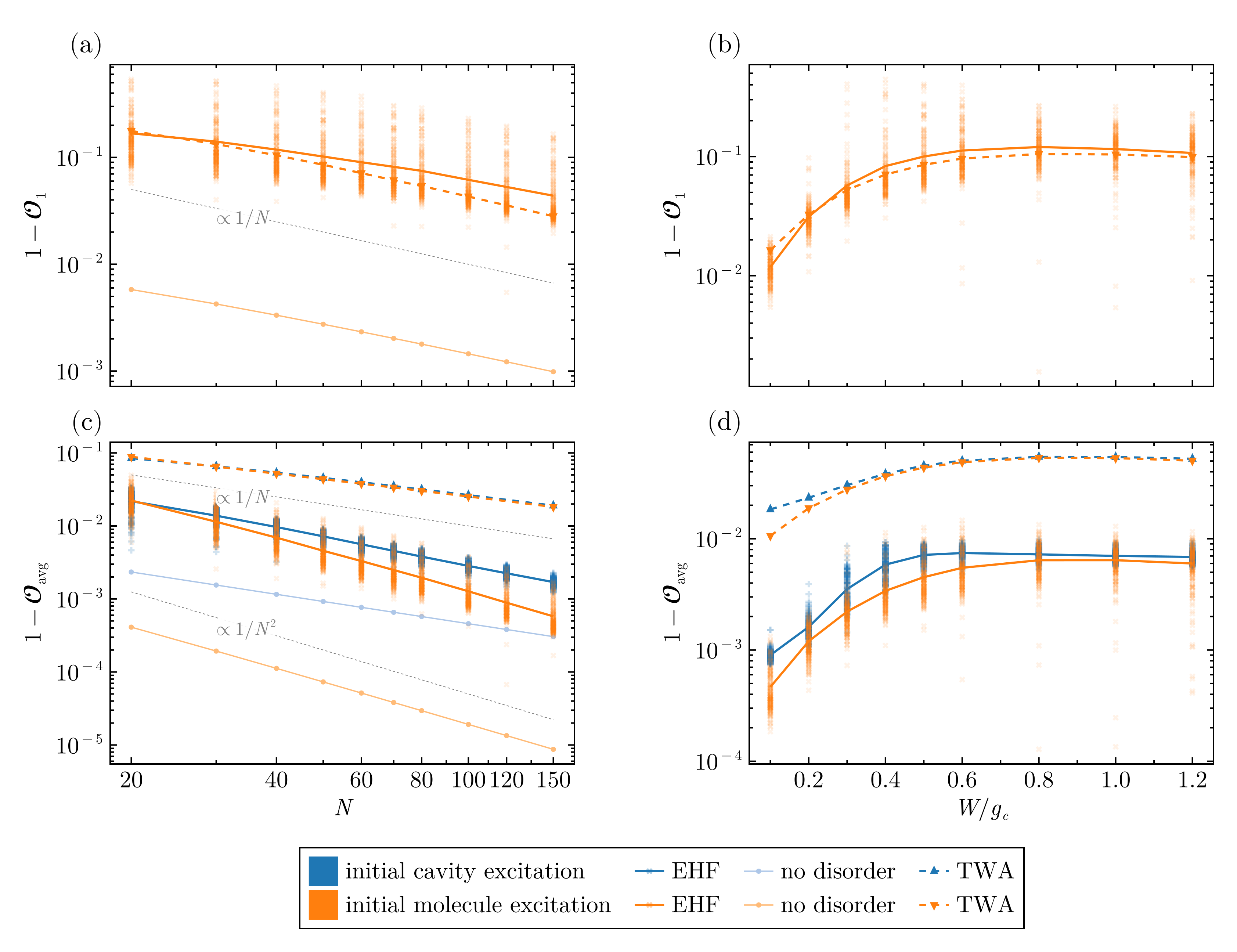}
    \caption{\textit{Testing semiclassical approximations after one vibrational period $t = 2\pi/\nu$ --} Plots show ``infidelities'' defined as $1 - \mathcal{O}_{1/\rm avg}$ with $\mathcal{O}_{1/\rm avg}$ the Wigner function overlap corresponding to the states of the single excited molecule $\mathcal{O}_{1}$, and the ensemble-averaged state $\mathcal{O}_{\rm avg}$, respectively. (a,b) Scaling of $1-\mathcal{O}_{1}$ as a function of molecule number $N$ for a fixed disorder strength of $W = 0.5 g_c$. (b) $1-\mathcal{O}_{1}$ as a function of disorder strengths for a fixed number of $N=50$ molecules.
    (c,d) The same plots for the ensemble-averaged infidelities $1-\mathcal{O}_{\rm avg}$. Shown are results for both initial cavity (blue) and molecular (orange) excitation scenarios. Markers: individual disorder realizations; solid lines: disorder averages. Light lines with circular markers: disorder-free case. Dotted gray lines indicate power laws (double-logarithmic scale) as guide to the eye. Parameters:  $N_D=200$ disorder realizations, $\nu = 0.3 g_c$, $\lambda=0.4$. Exact MPS results are converged with $\chi\leq 128$ and $n_{\rm max}^{\rm vib} = 8$.}
    \label{fig:overlap-scaling}
\end{figure*}

\section{Testing semiclassical approximations}
\label{sec:semiclassical}

In this section we analyze to what extent common semiclassical approaches capture the non-Gaussian features identified above. In particular, we focus on two approaches: i) the Ehrenfest mean-field method, which constrains vibrational modes to a product of coherent states, factorized from electronic and photonic degrees of freedom [see Eq.~\eqref{eq:ehf}]; and ii) the Truncated Wigner Approximation (TWA), which simulates dynamics classically, but averages over trajectories sampled from the initial Wigner distribution, thereby partially incorporating quantum fluctuations. Details on both methods are given in Appendix~\ref{sec:methods}. In polaritonic chemistry, they were previously used to compute quantum dynamics in several contexts~\cite{ehrenfestref1, ehrenfestref2, GDTWAVendrell, bond_TWA_NGS,kanakati_benchmarking_2026}.

\smallskip

To quantify how well those methods reproduce the exact quantum dynamics obtained from MPS simulations, we compute the Wigner function overlap of the vibrational states after one oscillation period $t = 2\pi/\nu$, computed from Eq.~\eqref{eq:wigner}. Denoting the Wigner functions obtained from our MPS simulations as $W^{1/\rm avg}_\text{MPS}(x,p)$ and the ones obtained from the Ehrenfest or TWA simulations as $W^{1/\rm avg}_\text{approx}(x,p)$, we compute
\begin{align}
    \mathcal{O}_{1/\rm avg} \equiv 2\pi\hbar \iint \! dx \, dp \; W^{1/\rm avg}_\text{MPS}(x,p) \, 
    W^{1/\rm avg}_\text{approx}(x,p)\label{eq:wigner-overlap}.
\end{align}
It is straightforward to show that this overlap corresponds to the Hilbert-Schmidt inner product between the respective reduced vibrational density matrices: $\mathcal{O}_{1/\rm avg}=\tr\left( \hat\xi_{1/\rm avg}^\text{MPS} \hat\xi_{1/\rm avg}^\text{approx}\right)$.

In Fig.~\ref{fig:overlap-scaling} we analyze the overlaps, again contrasting our different initial excitation scenarios and whether the vibrational states correspond to a single molecule ($\mathcal{O}_1$) or an ensemble average ($\mathcal{O}_{\rm avg}$). In particular, we analyze the scaling of the infidelity $1-\mathcal{O}_{1/\rm avg}$ as a function of the molecule number $N$ and the disorder strength $W$. As a function of the molecule number $N$, we always observe a decrease of the infidelity $1-\mathcal{O}_{1/\rm avg}$, i.e.,~an enhanced accuracy of the semiclassical approaches, in all scenarios. In all cases this decrease follows approximately a power law (double logarithmic scale). Strikingly, without disorder the infidelities are always significantly lower than with disorder. 

In Fig.~\ref{fig:overlap-scaling}(a) we focus on the modification of the vibrational state of a single molecule after its excitation. We find that without disorder, the Ehrenfest ansatz provides an excellent approximation with infidelities reaching to the $1-\mathcal{O}_1 \sim 10^{-3}$ level for $N \sim 100$ molecules. The presence of disorder has a drastic impact with a disorder strength of $W = 0.5g_c$ increasing the infidelities by almost two orders of magnitude. For large $N$ this is in line with the disorder-induced non-Gaussianity increasing with disorder in Fig.~\ref{fig:delta-scaling}(a). For our TWA simulations (dashed) line we find that they provide very similar results for the disordered scenario.

Remarkably, here we observe that for the considered range of molecule numbers, the power-law scaling of the infidelity $1-\mathcal{O}_1$ does not change significantly in the presence of disorder. This is in contrast to our findings for the nearly constant non-Gaussianity scaling in $\delta[\hat \xi_{1}]$ in Fig.~\ref{fig:delta-scaling}(a). Since fundamentally the Ehrenfest ansatz cannot feature finite non-Gaussianity, these results imply that in the thermodynamic limit either the non-Gaussianity must vanish, or the overlap scaling in Fig.~\ref{fig:overlap-scaling}(a) must turn to a constant scaling for $N \gtrsim 150$, outside the reach of our simulation capabilities. We conclude that the improving overlap at large $N$ should not be mistaken for intrinsic accuracy of the semiclassical methods. Significant non-Gaussian features can still be missed by the semiclassical assumptions, also for $N \gtrsim 100$.

\smallskip

As expected, for the ensemble-averaged scenario in Fig.~\ref{fig:overlap-scaling}(c) we again observe a faster vanishing infidelity: $1-\mathcal{O}_{\rm avg} \sim N^{-2}$ for an initial molecule excitation, and $1-\mathcal{O}_{\rm avg}$ for the initial cavity excitation. Those results are consistent with the scaling of the decreasing non-Gaussianity in Fig.~\ref{fig:delta-scaling}(c). For example, for systems with $N \sim 100$ molecules, Ehrenfest already reproduces exact vibrational states with an infidelity of $\sim 10^{-5}$.

\smallskip

In Fig.~\ref{fig:overlap-scaling}(b) and (d) we analyze the dependence of the infidelity as a function of the disorder strength $W$ for a system with $N=50$ molecules, for the initially excited molecule and the ensemble-averaged density matrices, respectively. For disorder strengths clearly within the strong coupling regime $0.1 g_c \leq W \leq 0.5 g_c $ we observe an increase of the infidelity with $W$, which implies that the semiclassical approximation becomes consistently worse with increasing disorder. We understand this as a direct consequence of the disorder-induced  entanglement~\cite{Wellnitz2022}, which similarly decreases only for values of $W \gtrsim 0.5 g_c$~\cite{Wellnitz2022}. For the single molecule predictions in Fig.~\ref{fig:overlap-scaling}(b) we find the TWA again to give quantitatively very similar results as the Ehrenfest approximation. For ensemble-averaged density matrices in  Fig.~\ref{fig:overlap-scaling}(d), however, we find that TWA results can lead to infidelities that are an order of magnitude larger than those obtained from Ehrenfest. Our results suggest that the Ehrenfest ansatz is more adapted to the problem in general than the TWA. However, we note that TWA enables very large scale simulations $N>10000$, inaccessible to the other techniques.

\section{Conclusion \& Outlook}
\label{sec:concl}

We have analyzed the modification of the nuclear vibrational dynamics due to the presence of a collective cavity coupling in competition with static local disorder in a toy model for polaritonic photo-chemistry. Specifically we studied exact dynamics in a disordered HTC model after incoherent photo-excitation of a single molecule or of the cavity mode. Confirming previous findings~\cite{Wellnitz2022} we observed that only in the presence of large disorder, the nuclear dynamics on the single-molecule level can be significantly modified by the cavity. With disorder such modifications remain robust also in large systems with many molecules ($N\gtrsim 100$) coupled to a single mode. We find modifications to be largest typically on the edge of the strong-coupling regime, and to be persistent also outside the strong-coupling regime $W \gtrsim g_c$.

Specifically, we have shown that the static energetic disorder leads to non-Gaussian features in the reduced vibrational states of the HTC model (Fig.~\ref{fig:somefeatures} and Fig.~\ref{fig:delta-scaling}). The single-molecule non-Gaussianity remains nearly constant over the numerically accessible range up to $N=150$, though in the absence of disorder it decreases as $\propto 1/N$ . In contrast, ensemble-averaged non-Gaussianity is always suppressed with increasing system size; however, for an initial cavity excitation, the decrease can be as slow as $1/N$, implying that the total (non-site-averaged) non-Gaussianity remains finite, consistent with the robustness of asymmetric nuclear distributions reported in Ref.~\cite{Wellnitz2022}. We therefore report that microscopic quantum signatures can persist at the edge of the collective strong coupling regime, even for large ensembles, provided the disorder is sufficiently strong.

Additionally, we have shown that the vibrational dynamics does not exhibit signatures of quantum thermalization on the ultrafast time scale (Fig.~\ref{fig:thermal}): the vibrational distributions do not approach a Boltzmann form, whether examined for individual realizations or averaged over disorder. Increasing disorder makes the distributions appear more thermal, but with a potential effective temperature being inconsistent with the post-excitation energy.

Moreover, comparing to exact calculations, we have found that common semiclassical methods fail quantitatively and qualitatively (Fig.~\ref{fig:overlap-scaling}). The disorder-induced non-Gaussianity is absent in such methods. The accuracy of both Ehrenfest and TWA methods strongly increases with the system size in the absence of disorder. However, large disorder causes differences to remain significant in our numerically accessible range (up to $N=150$). Our findings imply that quantum correlations that neither Ehrenfest mean-field nor TWA calculations can capture can remain significant also for large $N\sim 100$. This has implications for estimating the validity of semiclassical assumptions in ab-initio approaches for polaritonic chemistry with more realistic molecular models~\cite{bauman_perspective_2025}.

In the future, it would be important to analyze the impact of disorder also in more realistic models, going e.g.,~beyond the HTC model assumption and treating the nuclear dynamics beyond the harmonic approximation. It will then be crucial to answer to what extent the non-classical features observed here can have a real impact on photo-chemical reaction rates. Furthermore, it would be interesting to go beyond the unitary ultrafast regime, and to extend the discussion to steady states of driven-dissipative scenarios, which will then require to take the fast non-radiative vibrational relaxation dynamics into account. For the simplified disordered HTC model, it would be interesting to develop a further analytical understanding, also beyond the perturbative regime, explaining e.g., the scaling of the maximally achievable non-Gaussianities and respectively needed disorder strengths.

\begin{acknowledgements}

 We thank David Wellnitz, Liam Bond, Thomas Botzung, and Jérôme Dubail for helpful discussions. This work has been supported by the ERC Consolidator project MATHLOCCA (Grant nr.~101170485). R.~S.~is supported by the ML4Q program that has received funding from the European Union’s Horizon Europe research and innovation programme under the Marie Skłodowska-Curie grant agreement number 101120240. This work has received financing from the Interdisciplinary Thematic Institute QMat, as part of the ITI 2021-2028 program of the University of Strasbourg, CNRS and Inserm, and was supported by IdEx Unistra (ANR-10-IDEX-0002), and by SFRI STRAT’US project (ANR-20-SFRI-0012) and EUR QMAT ANR-17-EURE-0024 under the framework of the French Investments for the Future Program. Further support has been provided by the French National Research Agency under the France 2030 program ANR-23-PETQ-0002 (PEPR project QUTISYM) and project ANR-21-ESRE-0032 (PEPR project aQCess). Computations  were  carried  out  using  resources  of  the High Performance Computing Center of the University of Strasbourg, funded by Equip@Meso (as part of the Investments for the Future Program) and CPER Alsacalcul/Big Data.

\end{acknowledgements}

\appendix

\section{Gaussian reference states and entropy from covariance matrices}
\label{appendix-tau-closest-gaussian}
Here we outline the explicit construction of the closest Gaussian reference state $\hat \tau$ and the computation of its von Neumann entropy used to quantify non-Gaussianity. 
For a single harmonic oscillator mode, we define the quadrature operators $\hat R=(\hat x,\hat p)$, and define the covariance matrix elements of the state $\hat \rho$ as
\begin{equation}
    V_{ij} = \frac{1}{2}\left( \langle \hat R_i \hat R_j + \hat R_j \hat R_i\rangle - \langle \hat R_i\rangle \langle \hat R_j \rangle   \right).
\end{equation}
Although a Gaussian harmonic oscillator state is fully characterized by its first and second moments, there is no simple closed-form expression for the corresponding density matrix in the Fock basis as a direct function of the covariance matrix  $\hat V$.
Exceptions where simple expressions arise would for example be thermal states, which remain diagonal in the Fock basis. Thus, it is common to compute the von Neumann entropy of the state $\hat \tau$ directly from its covariance matrix \cite{Weedbrook2012,demarie2012pedagogicalintroductionentropyentanglement,juarez_quantum_2023}. This is done by expressing the entropy in terms of the symplectic eigenvalues $v$ of $\hat V$ \cite{AlessioSerafini2004}. 
The symplectic eigenvalues of the covariance matrix $\hat V$ can be found using Williamson's decomposition of the covariance matrix 
$\hat V = \hat S \hat D \hat S^T$ \cite{Eisert_2010}.
Here, $\hat D$ is the diagonal matrix ${\rm{diag}}(v,v)$ and $v$ the symplectic eigenvalue of the covariance matrix.
They can be obtained by computing 
\begin{equation}
    {\rm{det}}\left(\hat V \right) = {\rm{det}}\left(\hat S \hat D \hat S^T \right) = v^2 \ {\rm{det}}\left( \hat S  \hat S^T \right) = v^2,
\end{equation}
leading to $v = \sqrt{ \rm{det}\left( \hat V \right)}$. 
For a single-mode system the entropy is then given by
\begin{equation}
    S(\hat \tau) = \left( v + \frac{1}{2} \right){\rm{ln}}\left( v + \frac{1}{2} \right) -\left(v- \frac{1}{2} \right){\rm{ln}}\left( v - \frac{1}{2} \right).
\end{equation}
\section{Numerical methods} \label{sec:methods}
\subsection{Matrix product states} 
\label{ssec:MPS}
MPS are a well established numerical technique for 
computing the quantum dynamics of weakly entangled one-dimensional many-body systems \cite{vidal_efficient_2004,paeckel_timeevolution_2019}. Their efficiency comes from a local, low-rank approximation of the full state, constructed by truncating the Schmidt decomposition across each bond to a large-enough bond dimension $\chi$. We evolve the state using second order Time-Evolved Block Decimation, based on the usual Trotter decomposition~\cite{sornborger1999higher} of the time-evolution operator of our Hamiltonian~\cite{paeckel_timeevolution_2019}. To handle the all-to-all coupling introduced by the cavity mode, we swap the cavity site through the chain for each Trotter sweep. We make sure that all MPS results are converged in terms of the time-step size and in terms of the bond dimensions $\chi$ quoted in the figure captions.

\subsection{Ehrenfest mean-field}
\label{ssec:EHF}
By ``Ehrenfest mean-field'' we denote dynamics obtained using a product-state ansatz that factorizes the vibrational degrees of freedom from the rest:
\begin{align}
    \ket{\psi^\text{EHF}(t)} = \ket{\phi(t)} \otimes \prod_i \mathcal{D}\left(\lambda_i(t), \hat{b}_i\right) \ket{0}_\text{vib,i}, \label{eq:ehf}
\end{align}
where $\mathcal{D}(\lambda, \hat{b}_i) \equiv \exp[\lambda(\hat{b}_i^\dagger - \hat{b_i})]$ denotes the displacement operator. In the disorder-free case ($W=0$) this ansatz yields an analytically tractable problem: writing $ \ket{\phi(t)}$ as a superposition of polaron eigenstates of the Tavis-Cummings model leads to phenomena such as polaron decoupling \cite{herrera_cavitycontrolled_2016,zeb_exact_2018}. For the disordered case studied here, we obtain Ehrenfest dynamics numerically within the MPS framework by carefully restricting the corresponding bond dimension to $\chi=1$, which enforces the product-state structure in Eq.~\eqref{eq:ehf} throughout the evolution.

\subsection{Truncated Wigner approximation}
\label{ssec:TWA}
The Truncated Wigner Approximation (TWA) \cite{polkovnikov_2010} is a semiclassical numerical scheme that is often used to compute the quantum dynamics of many-body systems. This method is based on the phase space formalism of quantum mechanics where Hilbert space operators $\hat O$ are mapped through the Weyl-Wigner transform to Weyl symbols $O_W$, functions of classical phase space variables $(\mathbf{x},\mathbf{p})$. In the TWA, each Weyl symbol $O_W$ evolves according to the classical equation of motion $\dot O_W(\mathbf{x},\mathbf{p}) = \{O_W(\mathbf{x},\mathbf{p}),H_W(\mathbf{x},\mathbf{p})\}$, where $\{.,.\}$ denotes the Poisson bracket and $H_W$ is the Weyl symbol of the Hamiltonian of the system. The Poisson bracket ensures that any Weyl symbol factorizes into phase space variables such that we only need the evolution of $(\mathbf{x}(t),\mathbf{p}(t))$ to describe the system completely. This makes the TWA an efficient numerical method in the case of an initial positive Wigner function: the initial coordinates $(\mathbf{x}(0),\mathbf{p}(0))$ can be sampled from the initial Wigner function $W(\mathbf{x},\mathbf{p},t=0)$ and evolved with the classical equations of motion. Expectation values are then computed statistically with $\expval{\hat O}(t) = \frac{1}{n_{traj}}\sum_{i=1}^{n_{traj}}O_W(\mathbf{x}_i(t),\mathbf{p}_i(t))$ and Wigner functions are obtained by making a histogram of all the trajectories. 

For spin-1/2 systems, we introduce a discrete phase space \cite{wooters_1987} with phase space variables $(\mathbf{s}_x,\mathbf{s}_y,\mathbf{s}_z)$ defined as the Weyl symbols of the Pauli operators $(\hat{\boldsymbol{\sigma}}_x,\hat{\boldsymbol{\sigma}}_y,\hat{\boldsymbol{\sigma}}_z)$ and a numerical scheme based on the TWA, the Discrete Truncated Wigner Approximation (DTWA)~\cite{dtwa,gdtwa}. In the DTWA, phase space variables describe completely the system and evolve according to the classical equations of motion given by the generalized Poisson bracket \cite{generalizedpoissonbracket_1980}. The initial conditions are sampled according to a discrete Wigner function $W(\mathbf{s}_x,\mathbf{s}_y,\mathbf{s}_z)$ and the expectation values are again given by $\expval{\hat O}(t) = \frac{1}{n_{traj}}\sum_{i=1}^{n_{traj}}O_W(\mathbf{s}_x(t),\mathbf{s}_y(t),\mathbf{s}_z(t))$.

To compute the dynamics of the HTC model, we use a hybrid scheme based on the TWA and DTWA algorithms. The vibrational degrees of freedom ($\hat x_i = (\hat b_i + \hat b_i^\dagger)/\sqrt{2},~\hat p_i = -\textrm{i}(\hat b_i + \hat b_i^\dagger)/\sqrt{2}$) are treated using TWA and initially sampled with the Wigner function of the ground state of the harmonic oscillator. Cavity ($\hat a_x,~\hat a_y,~\hat a_z$) and electronic ($\hat \sigma_x,~\hat \sigma_y,~\hat \sigma_z$) degrees of freedom are treated using DTWA and are sampled with the Wigner function of a $\ket{\uparrow}$ or $\ket{\downarrow}$ state.
\\

\bibliography{main.bib}

\end{document}